\documentclass{aa} 

\usepackage{graphicx}
\usepackage[rightcaption]{sidecap}
\usepackage{txfonts}
\usepackage{hyperref}
\usepackage{xcolor}

\begin{document} 

    \title{Structural parameters, chronological age and dynamical age of the LMC globular cluster NGC~1754 \thanks{Based on observations with the
    NASA/ESA HST, obtained under programme GO 16361 (PI: Ferraro). The
    Space Telescope Science Institute is operated by AURA, Inc., under
    NASA contract NAS5-26555.}} 

   \author{Camilla Giusti
          \inst{1}
          \inst{2}
          \and
          Mario Cadelano
          \inst{1}
          \inst{2}
          \and
          Francesco R. Ferraro
          \inst{1}
          \inst{2}
          \and
          Barbara Lanzoni
          \inst{1}
          \inst{2}
          \and
          Cristina Pallanca
          \inst{1}
          \inst{2}
          \and
          Enrico Vesperini
          \inst{3}
          \and
          Emanuele Dalessandro
          \inst{2}
          }

\institute{Dipartimento di Fisica \& Astronomia, Universit\`a degli Studi di Bologna, via Gobetti 93/2, I-40129 Bologna, Italy\\
\and
INAF - Astrophysics and Space Science Observatory Bologna, Via Gobetti 93/3, 40129, Bologna, Italy \\
\and 
Dept. of Astronomy, Indiana University, Bloomington, IN 47401, USA\\
}


 
\abstract{In the context of a new systematic study of the properties of the most compact and massive star clusters in the Large Magellanic Cloud (LMC), here we present the determination of the chronological age, the structural parameters, and the dynamical age of NGC 1754. We used high-resolution images taken with the WFC3/HST instrument, both in optical and near-ultraviolet filters. The high quality of the images made it possible to construct the star density profile from resolved star counts, and to fit the observed profile with an appropriate King model to obtain the structural parameters (e.g. core, half-mass, and tidal radii). Our findings confirm that NGC 1754 is a very compact globular cluster with a core radius of only 0.84 pc. The analysis of the same dataset allowed us to confirm a very old age ($t=12.8\pm0.4$ Gyr) for this system, thus further consolidating the indication that the process of globular cluster formation started at the same cosmic time both in the LMC and in the Milky Way, independently of the characteristics of the host environment. We have also used the empirical method called ``dynamical clock'' to estimate the dynamical age of the system. This consists of quantifying the degree of central segregation of blue straggler stars (BSSs) using the $A^+_{rh}$ parameter, which is defined as the area enclosed between the cumulative radial distribution of BSSs and that of a reference (lighter) population. This method yielded a value of $A^+_{rh}=0.31\pm0.07$, which is the highest measured so far for LMC clusters, pointing to an advanced dynamical age for the cluster, possibly on the verge of core collapse. The results presented here for NGC 1754 confirm that the natural dynamical evolution of globular clusters plays a role in shaping the age-core radius distributions observed in the LMC.}


   \maketitle
%

\begin{figure*}
     \centering
         \includegraphics[scale = 0.48]{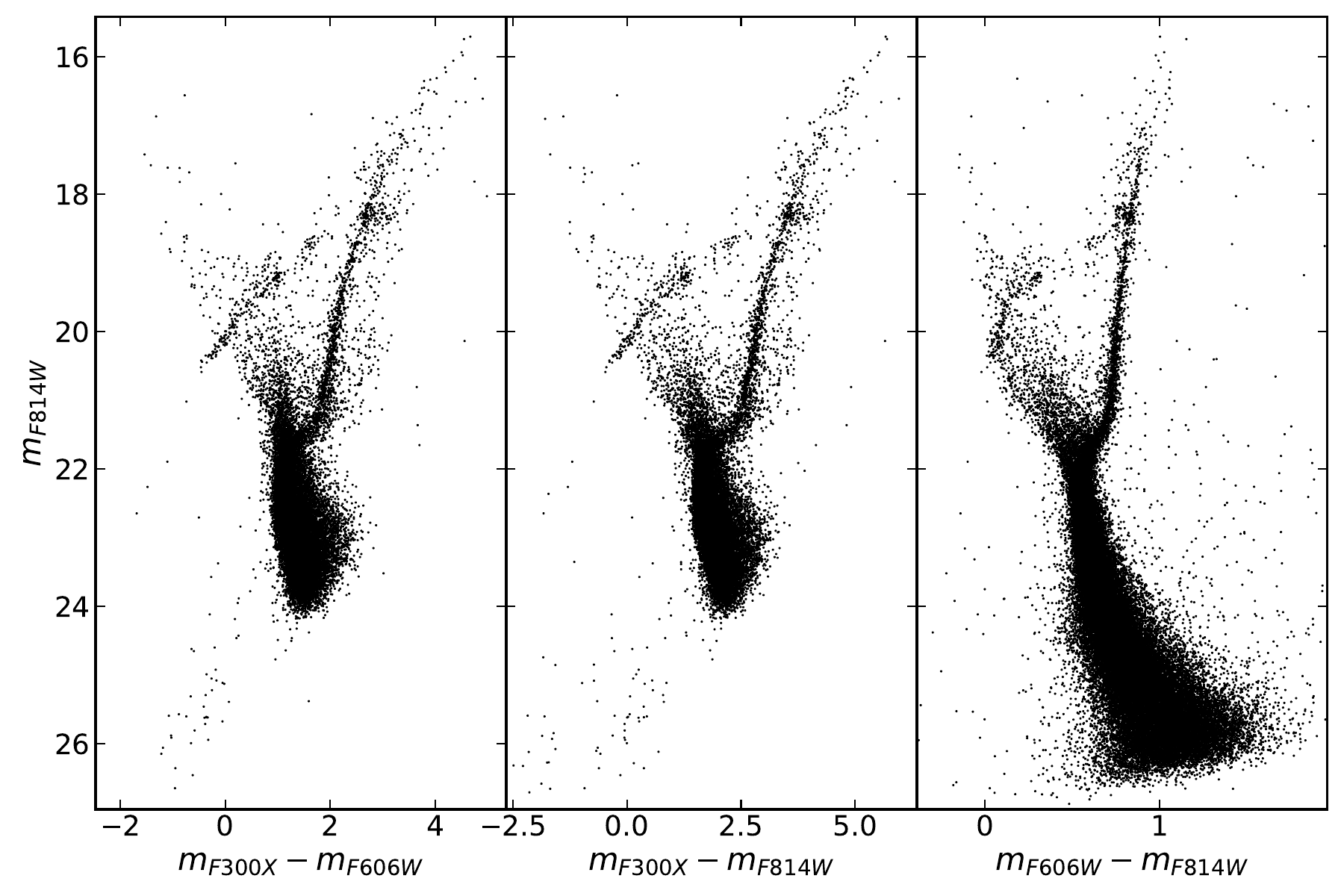}
         \caption{CMD of NGC 1754 obtained from the data reduction of the WFC3 dataset in all the filter combinations.}
         \label{cmds}
\end{figure*}

\section{Introduction}
\label{sec:intro}
At odds with the Milky Way (MW), the Large Magellanic Cloud (LMC) hosts a system of globular clusters (GCs) that covers a wide range of ages (from a few million to several billion years) and masses that vary between a few $10^3$ and several $10^5 M_{\odot}$ (see Fig.1 in \citealt{ferraro+2019}).  
This variety provides valuable insights into the processes of star cluster formation and evolution under conditions different from those in the MW \citep{olszewski+1996,olsen+1998, brocato+1996, mackey+2003, ferraro+2006, baumgardt+2013, cadelano+2022a, mucciarelli+2021}. Moreover, it offers a unique opportunity to study the major transitions phases in stellar evolution \citep{renzini+1986} across cosmic time \citep{ferraro+1995,ferraro+2004,mucciarelli+2006}

Previous studies \citep[see][]{elson+1989, elson+1991, mackey+2003} have revealed a peculiar behavior when analyzing the core radius of LMC GCs as a function of their chronological age (the so-called {\it size–age conundrum}): while the youngest systems ($t<3$ Gyr) all have compact core radii ($r_c < 2.5$ pc), older clusters span a range from 2.5 to 10 pc.
This unusual behavior has been explained as a sort of evolutionary sequence, with the younger clusters representing the precursors to the older ones. In particular, it was suggested that young (compact) clusters might expand over time due to gravitational interactions occurring in their centers and involving binary black holes \citep{mackey+2003, mackey+2008}. However, \citet{ferraro+2019} challenged this hypothesis, pointing out that the observed cluster distribution in mass and galactocentric distance does not support this evolutionary connection. 
Specifically, younger clusters cannot be representative of the young progenitors of older systems, because they are all less massive ($M < 10^5 M_{\odot}$) and located closer to the LMC's center ($R_g < 5$ kpc).

In this framework, an alternative explanation of the oldest cluster behavior (with no need to invoke the action of binary black holes) is that the observed spread in core radii is the natural result of internal dynamical aging, with the most compact systems being the most dynamically evolved. In fact, GCs are collisional systems where the gravitational interactions among stars progressively alter the initial conditions over time.  
The most massive stars tend to transfer kinetic energy to lower-mass objects and progressively sink toward the system's center (mass segregation); the energy transfers lead to the gradual escape of (preferentially low-mass) stars from the system (evaporation); in addition to the decrease in the mass of the cluster, the effects of two-body relaxation drive the evolution of the cluster's structure, yielding a progressive contraction of the core and an increase of the core density up to the so-called ``core collapse''. The timescales of this dynamical evolution depend in a complex way on the cluster's internal properties as well as on the strength of the host galaxy tidal field, thus differing from cluster to cluster. Hence, star clusters with the same chronological age can be in different stages of internal dynamical evolution.  
\citet{ferraro+2019} analyzed five LMC clusters with a core radius between 1 and 8 pc and with an age of $\sim$ 12-13 Gyr, and concluded that their spread in core radii is the result of dynamical aging, with the most compact systems being the most dynamically evolved (consistent with what found in MW GCs; see \citealp{ferraro+2018a,ferraro+2020,ferraro+2023, cadelano+2022b}). To estimate the dynamical age of star clusters, \citet{ferraro+2019} used the so-called ``dynamical clock'' \citep[see also][]{ferraro+2012,lanzoni+2016}, an empirical method based on the degree of central segregation of blue straggler stars (BSSs). Indeed, BSSs are ideal for this purpose since they are more massive (M$_{\rm BSS}\sim 1-1.4 M_\odot$; \citealt{fiorentino+2014, raso+2019}) than the average stars in old clusters ($m \sim 0.3-0.4 M_\odot$), which makes them sink to the bottom of the potential well under the action of dynamical friction. BSSs are also easily identifiable in a color-magnitude diagram (CMD), where they populate an extension of the main-sequence above the turn-off. The parameter used to quantify the BSS segregation is $A^+_{rh}$, defined as the area enclosed between the cumulative radial distribution of BSSs and that of a reference  population of lighter stars within one half-mass radius ($r_h$) from the cluster center \citep{alessandrini+2016,lanzoni+2016}. Empirical studies of MW GCs \citep{lanzoni+2016,ferraro+2018a,ferraro+2020,ferraro+2023} have demonstrated a strong correlation between the value of $A^+_{rh}$ and the central relaxation time, proving the validity of the method. 
 
However, to fully understand the behavior of the oldest clusters in the LMC it is necessary to extend the analysis of \citet{ferraro+2019} to the most compact systems, with $r_c<$1 pc, which were not included in that work.
The study started in \citet{giusti+2024b}, where we analyzed NGC 1835 (with an age of 12.5 Gyr and a core radius of 0.84 pc), finding that it has a very advanced dynamical age ($A^+_{rh}$ = 0.30). Similarly, in this work we take advantage of a new set of optical and UV high-resolution Hubble Space Telescope (HST) images to study the globular cluster NGC 1754. Literature data indicate that it has a very old chronological age ($t\sim 15$ Gyr;
\citealt{mackey+2003}), a particularly compact core radius ($r_c = 0.88$ pc;
\citealt{mackey+2003}) and a metallicity of [Fe/H]$\sim -1.45$ 
\citep{mucciarelli+2021}.

The paper is structured as follows. In Sect. \ref{sec:dataanalysis} we
describe the dataset, the data reduction process and we present the main features of the CMD.  Sect. \ref{sec:results} presents the main results of the paper: the determination of the cluster center; the measurements of the chronological age, distance modulus, reddening and metallicity; the construction of the density profile based on resolved star counts and the determination of the structural parameters from the fit of the profile with a King model; the measurements of the dynamical age through the dynamical clock method. Finally, in Sect. \ref{disc} we present the summary and discussion of the results.

 \begin{figure}
     \centering
         \includegraphics[scale = 0.48]{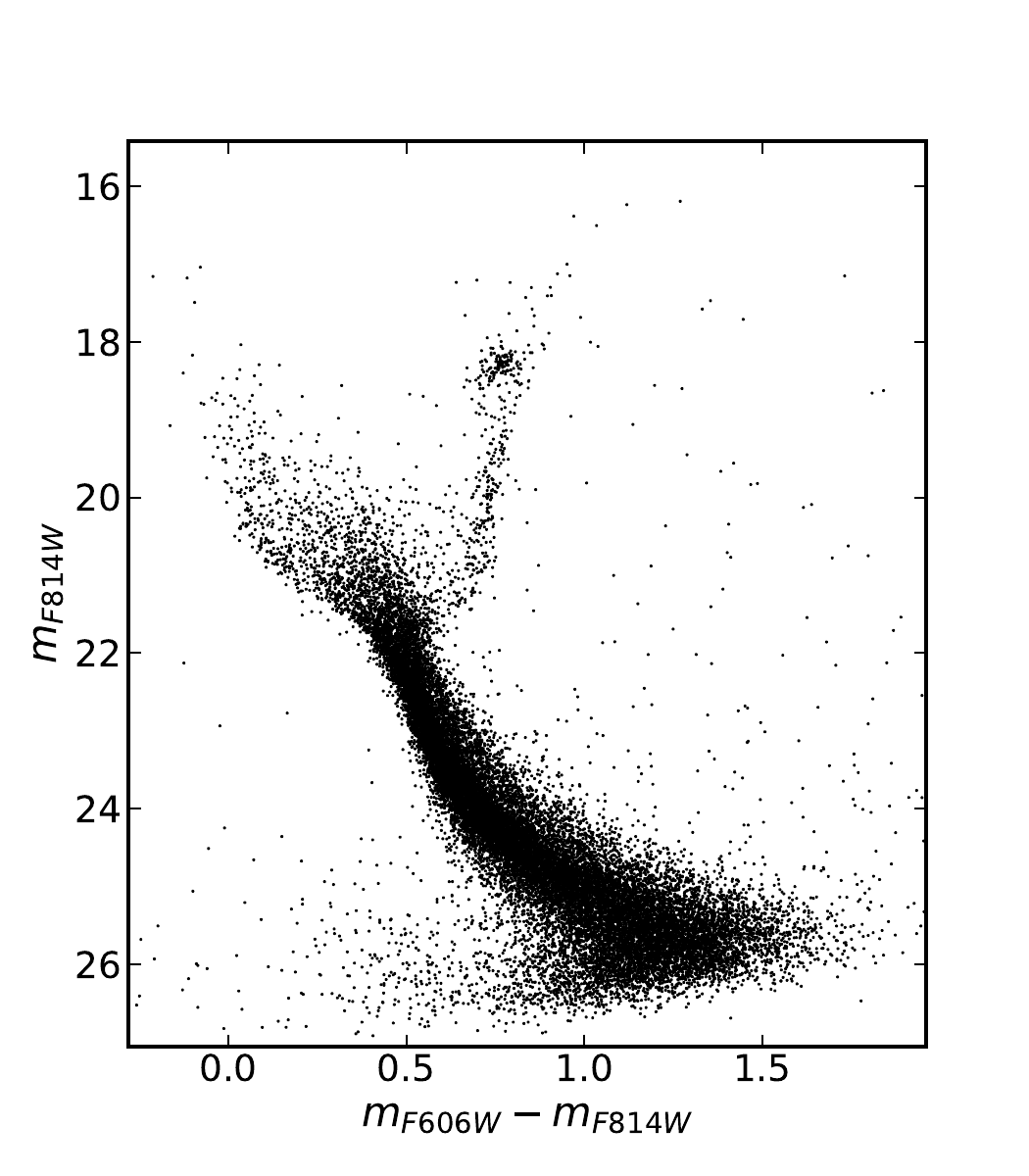}
         \caption{CMD of the field region obtained from the data reduction of the ACS dataset.}
         \label{cmdacs}
\end{figure}

\section{Data analysis}
\label{sec:dataanalysis}
For the photometric analysis of the cluster, we used a set of 16 high-resolution images taken by the HST's Wide Field Camera 3 (WFC3/HST) with the UVIS1 channel centered on the cluster. Specifically, six images ($3\times$ 927 s, $1\times$ 939 s, $1\times$ 940 s, $1\times$ 953 s) were acquired in the near-UV F300X filter, six ($5\times$ 414 s, $1\times$ 415 s) in the F606W, and four ($2\times$ 657 s, $2\times$ 700 s) in the F814W.  The use of the UV filter is crucial in this investigation since it offers optimal sensitivity to the detection of hot stars, as BSSs and stars on the hot horizontal branch (HB; see \citealt{ferraro+1997,ferraro+2001}). Indeed this approach has been very effective in the identification of the extreme hot tail in the HB of NGC 1835 \citep{giusti+2024a} and in the study of the BSS population in the same cluster \citep{giusti+2024b}.

The information from this dataset was complemented with a second one consisting of 13 images taken by the HST's Advanced Camera for Surveys (ACS) in optical filters (F606W and F814W) and sampling a field region located at $\sim5\arcmin$ from the center of the cluster. 
The data reduction of both datasets was performed using the DAOPHOT II software \citep{stetson+1987} with a procedure similar to that used in \citet[see also \citealt{cadelano+2017,onorato+2023, chen+2021, chen+2023}]{giusti+2024a, giusti+2024b}. We obtained the PSF model corresponding to each image in each filter by using approximately 200 bright, isolated, unsaturated stars as a reference. We performed a search for the sources at a level of approximately 5$\sigma$ from the background and fitted the PSF model to each of these sources. We then created a master-list including all the sources detected in at least half of the images of each chip, and we used the command ALLFRAME to force the fit of these stars to their corresponding positions in all the other images. This approach allowed us to retrieve previously missed hot stars (such as BSSs and white dwarfs), while at the same time recovering the faintest part of the main sequence.
For each identified star, the magnitudes estimated in different images were combined using DAOMATCH and DAOMASTER. In the end, we obtained a catalog with instrumental magnitudes and instrumental positions for $\sim$ 53400 sources in the WFC3 dataset, and $\sim$ 25300 stars in the ACS dataset. We then applied the standard calibration process to these sources, converting the magnitudes to the VEGAMAG reference system by applying the appropriate aperture corrections and the zero points found on the HST website. Finally, after correcting for geometric distortions using the coefficients of \citet{bellini+2011} for WFC3 and of \citet{meurer+2003} for ACS, the catalogs were astrometrized by cross-correlation with a \textit{Gaia} Data Release 3 catalog \citep{gaia+2023} of the same area, so that the coordinates have been transferred to the absolute coordinate system ($\alpha, \delta$).

In  Fig.\ref{cmds} and Fig.\ref{cmdacs} we present, respectively, the CMD of the cluster obtained from the reduction of the WFC3 dataset in all filter combinations, and the CMD of the field region from the ACS dataset. As expected, the cluster CMD exhibits some field contamination. Specifically, in the optical CMDs the main sequence (MS) and the red giant branch (RGB) of the field nearly overlap with those of the cluster. In contrast, in the CMDs using the F300X filter, the RGBs of the field and the cluster are well separated. Despite field contamination, the cluster sequences remain clearly visible, with a main sequence ranging down to magnitude $m_{\rm F814W} \sim 26$ and the turn-off located at $m_{\rm F814W} \sim 21.5$. The HB is distinguishable in the magnitude range $18.5 \leq m_{\rm F814W} \leq 21$, crossing the extended blue plume of the field. In particular, along the HB in the $m_{\rm F814W},m_{\rm F606W}- m_{\rm F814W}$ CMD it is possible to appreciate a group of stars at redder colors ($0.6 \leq (m_{\rm F606W}-m_{\rm F814W}) \leq 0.8$) and another one at bluer colors ($0 \leq (m_{\rm F606W}-m_{\rm F814W}) \leq 0.4$), separated by the instability strip region.

\begin{figure}
     \centering
         \includegraphics[scale = 0.48]{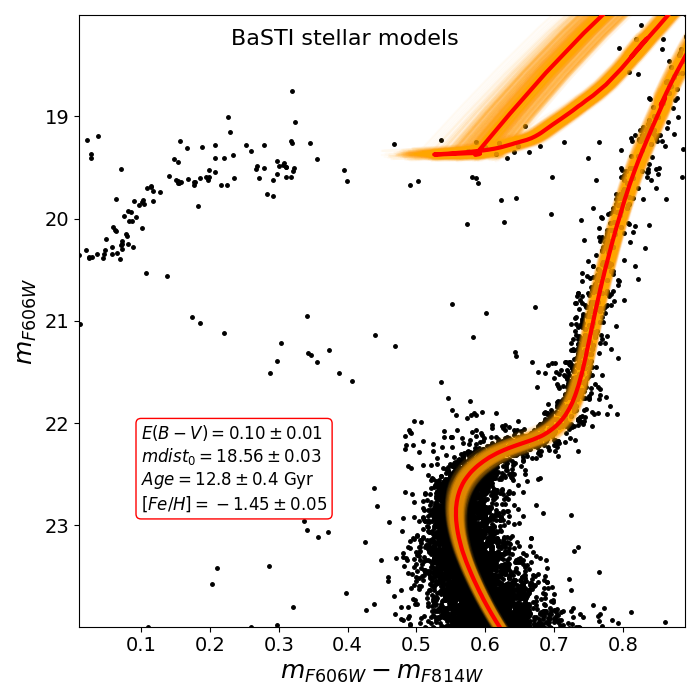}
         \caption{Field-decontaminated CMD of NGC 1754 in the radial annulus between 5\arcsec and 30\arcsec (black dots),  with the best-fit BASTI isochrone superposed as a red solid line. The orange-shaded envelope represents the $1\sigma$ uncertainty region of the best-fit isochrone. 
         The best-fit values for the reddening $E(B-V)$, distance modulus $(m-M)_0$, age ($t$), and metallicity ([Fe/H]) are also labelled.}
         \label{age}
\end{figure}

\section{Results}
\label{sec:results}
\subsection{Gravitational center}
\label{sec:center}
We re-evaluated the position of the gravitational center of the cluster through an iterative procedure first described in \citet[see also \citealt{lanzoni+2007, lanzoni+2019, giusti+2024b}]{montegriffo+1995}. The high spatial resolution of the images allowed the determination of the center by averaging the positions (right ascension and declination) of the stars located within a fixed distance $d$ from a first-guess value, and selected brighter than a certain magnitude limit. This approach helps to minimize biases that may arise when identifying the gravitational center with the location of the brightness peak, especially in the presence of very bright stars. Specifically, the procedure starts with a first-guess estimate of the center (in our case, the literature value provided by \citealt{mackey+2003}, $\alpha = 04^h:54^m:18^s.9$, $\delta=
-70^{\circ}:26\arcmin:31\arcsec$). 
The selection distance $d$ must exceed the cluster's core radius ($r_c$ =3.61\arcsec, \citealt{mackey+2003}) to avoid operating in a region of nearly constant density, which would hinder convergence. At the same time, it must remain sufficiently small to avoid losing sensitivity to the central area. Similarly, the magnitude limits are chosen as a compromise: they must be faint enough to ensure high statistical significance, while avoiding incompleteness issues. The averaged positions of the selected sample of stars yield a new center, which serves as the starting point for the next iteration. Convergence is achieved when the positions of two consecutive centers differ by less than 0.01\arcsec. We repeated the procedure for three values of maximum distance from the center ($d< 8,12,16$\arcsec) and three magnitude limits ($m_{\rm F814W} \leq 21.5,22.0,22.5$), thus providing nine samples of stars, each with a minimum of a few hundred objects and a completeness level above 60\% (see Sect.\ref{BSS}). As expected, each of the nine center calculations resulted in small differences, influenced by perturbations in the star distribution and reflecting the natural variations within the cluster's core region. For this reason, we chose as value of the gravitational center of NGC 1754 the average of the nine determinations, namely $\alpha = 04^h:54^m:18.6549^s$, $\delta=
-70^{\circ}:26\arcmin:31.0329\arcsec$, and their standard deviation $\sigma = 0.2$ as the error. The center thus found is only $ d\sim 1.2$\arcsec from the center of the literature.

\subsection{Chronological age, reddening and distance modulus}
\label{sec:age}
We determined the chronological age of the cluster using a Bayesian method similar to the one adopted in previous studies (see e.g. \citealp{cadelano+2020b,deras+2023,deras+2024}). The procedure consists of an isochrone fitting technique that compares the observed CMD of the cluster with a set of isochrones, exploring reasonable grids of values for the relevant parameters and allowing the simultaneous estimate of the chronological age, reddening, distance modulus and metallicity. 
We restricted our analysis to the CMD in a radial annulus between 5\arcsec and 30\arcsec, to avoid the central region where photometric errors are larger due to severe crowding and the outer region where field stars begin to have strong importance. We further reduced field star contamination by applying a statistical decontamination process described also in \citet{dalessandro+2019,giusti+2023, giusti+2024a}. This procedure consists of comparing the cluster CMD (in this case the radial one between 5\arcsec and 30\arcsec) with a field CMD (from the ACS dataset) in a region of equivalent area, and removing from the former the likely field interlopers. First of all, for each object present in the field CMD, we identified the closest star in the cluster CMD, with the reciprocal distance measured as the square root of the magnitude and color differences summed in quadrature. Then, we classified as possible field intruders and removed from the cluster CMD all the selected stars with reciprocal distance smaller than 0.30 magnitudes. At the end of the cleaning process the key evolutionary sequences (RGB, sub-giant branch and the MS turnoff region) of the cluster population are clear and well defined in the CMD (see black dots in Fig.\ref{age}). 

We focused the comparison between the CMD and the isochrones in the regions around the MS turn-off, sub-giant branch, and the lower portion of the RGB (21.3$<m_{\rm F606W}<$23), as these are more sensitive to age and metallicity variations.  We used BASTI isochrones \citep{hidalgo+2018,pietrinferni+2021} downloaded with standard helium abundance ($Y = 0.25$) and [$\alpha$/Fe] = +0.4, for a wide range of ages (from 9.0 to 14.0 Gyr, with a step of 100 Myr) and metallicities (from [Fe/H] = $- 1.8$ to [Fe/H] = $- 0.95$, in steps of 0.05). 
The comparison between the field-decontaminated CMD and the isochrones has been performed through a Markov Chain Monte Carlo (MCMC) sampling technique and we used the {\it emcee} code to sample the posterior probability distribution in the n-dimensional parameter space defined by age, distance modulus, reddening and metallicity \citep{foreman+2013,foreman+2019}. For the metallicity we assumed a Gaussian prior distribution with peak and dispersion values [Fe/H]$= -1.45 \pm 0.05$ \citep{mucciarelli+2021}. Gaussian priors were also adopted for the reddening and the distance modulus, assuming as Gaussian mean and dispersion the values determined by aligning the CMDs of M3 and M13 with that of NGC 1754. The former are two twin clusters in our galaxy with a metallicity similar to NGC 1754 and well-defined parameters: $(m-M)_0 = 15.00\pm 0.04$ and $E(B-V)=0.01$ for M3, $(m-M)_0 = 14.32\pm 0.05$ and $E(B-V)= 0.02$ for M13 \citep[see e.g.][]{dalessandro+2013a}. From the color and magnitude shifts needed to superpose their CMDs to that of NGC 1754, we found for the latter a color excess $E(B - V) = 0.13 \pm 0.02$ and a distance modulus $(m - M)_0$ =18.58 $\pm$ 0.05, which have been used as Gaussian priors in the MCMC procedure. When converting absolute magnitudes of the isochrones to the observed frame, we used the extinction coefficients $R_{\rm F606W}=2.8192$ and $R_{\rm F814W}=1.8552$ \citep{cardelli+1989, odonnell+1994}. 

Fig.\ref{age} shows the best-fit isochrone (solid red line) with its 1$\sigma$ uncertainty (orange shaded envelope) superposed to the field-decontaminated CMD of NGC 1754 (black dots). The best-fit values obtained with this procedure are an age t = 12.8 $\pm$ 0.4 Gyr, a reddening $E(B-V) = 0.10 \pm 0.01$, a distance modulus $(m-M)_0$  = 18.56 $\pm$ 0.03, and a metallicity [Fe/H] = $-1.45 \pm 0.05$. The uncertainties are determined using the MCMC procedure, by calculating the 16$^{th}$ and 84$^{th}$ percentiles of each parameter's probability distribution. The reddening and distance modulus obtained in this work are compatible, within the errors, with the literature values quoted in \citet{olsen+1998}, namely $E(B-V) = 0.09 \pm 0.02$ and $(m-M)_{0}$ ranging between $18.60 \pm 0.16$ and $18.70 \pm 0.16$. In addition, the metallicity value is fully consistent with that measured from spectroscopy by \citet{mucciarelli+2021}.

\begin{figure}
     \centering
         \includegraphics[scale = 0.35]{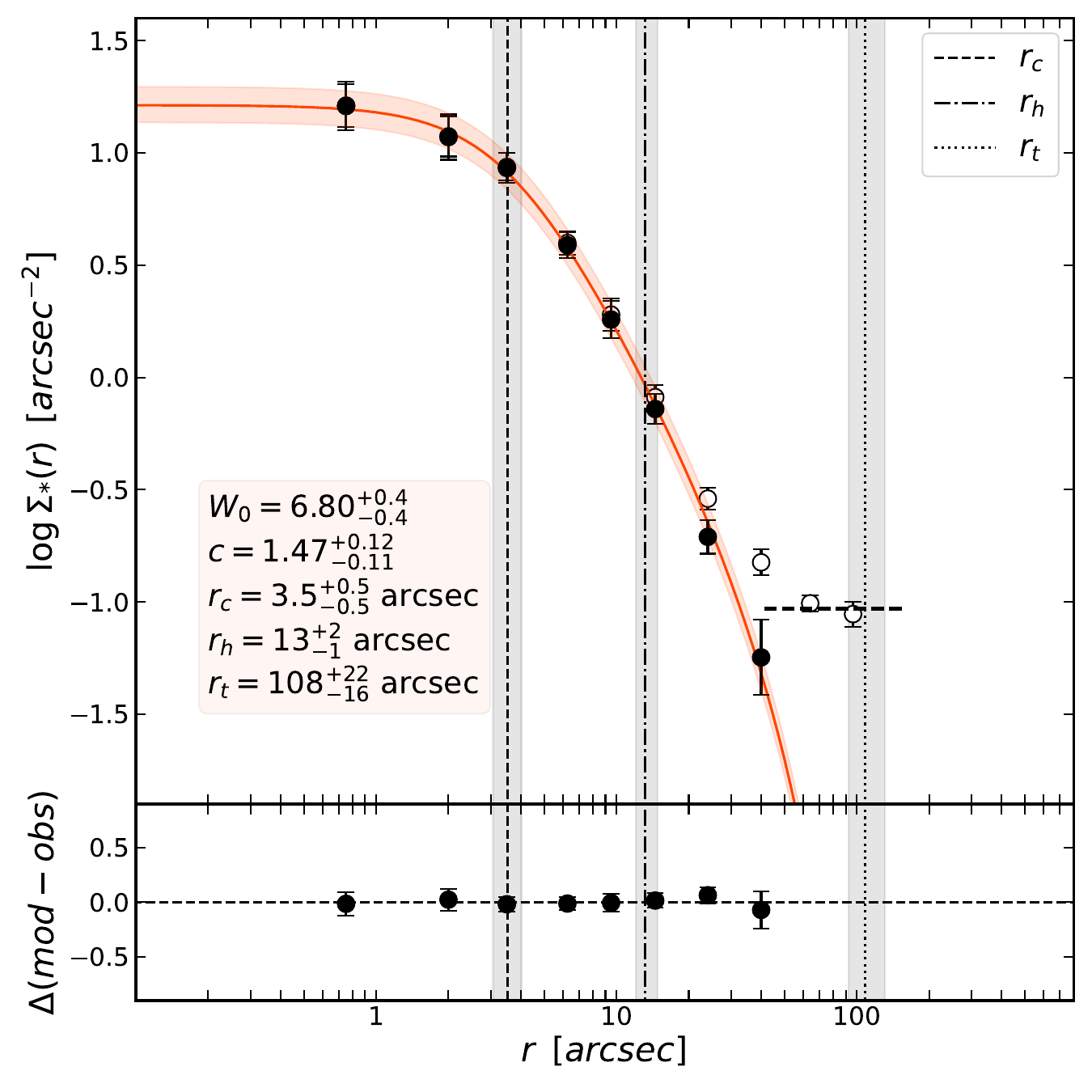}
         \caption{The projected density profile of NGC 1754, derived from resolved star counts (empty circles). The horizontal dashed
           line represents the mean LMC field density, which was subtracted to obtain the background-subtracted profile (filled circles). The solid red line illustrates the best-fit King model, while the red stripe indicates the $\pm 1\sigma$ range of solutions.  Vertical lines denote the core radius (dashed line), half-mass radius (dot-dashed line), and tidal radius (dotted line), with their respective 1$\sigma$ uncertainties highlighted by grey stripes.}
\label{profili}
\end{figure}

\begin{figure*}
     \centering
         \includegraphics[scale = 0.2]{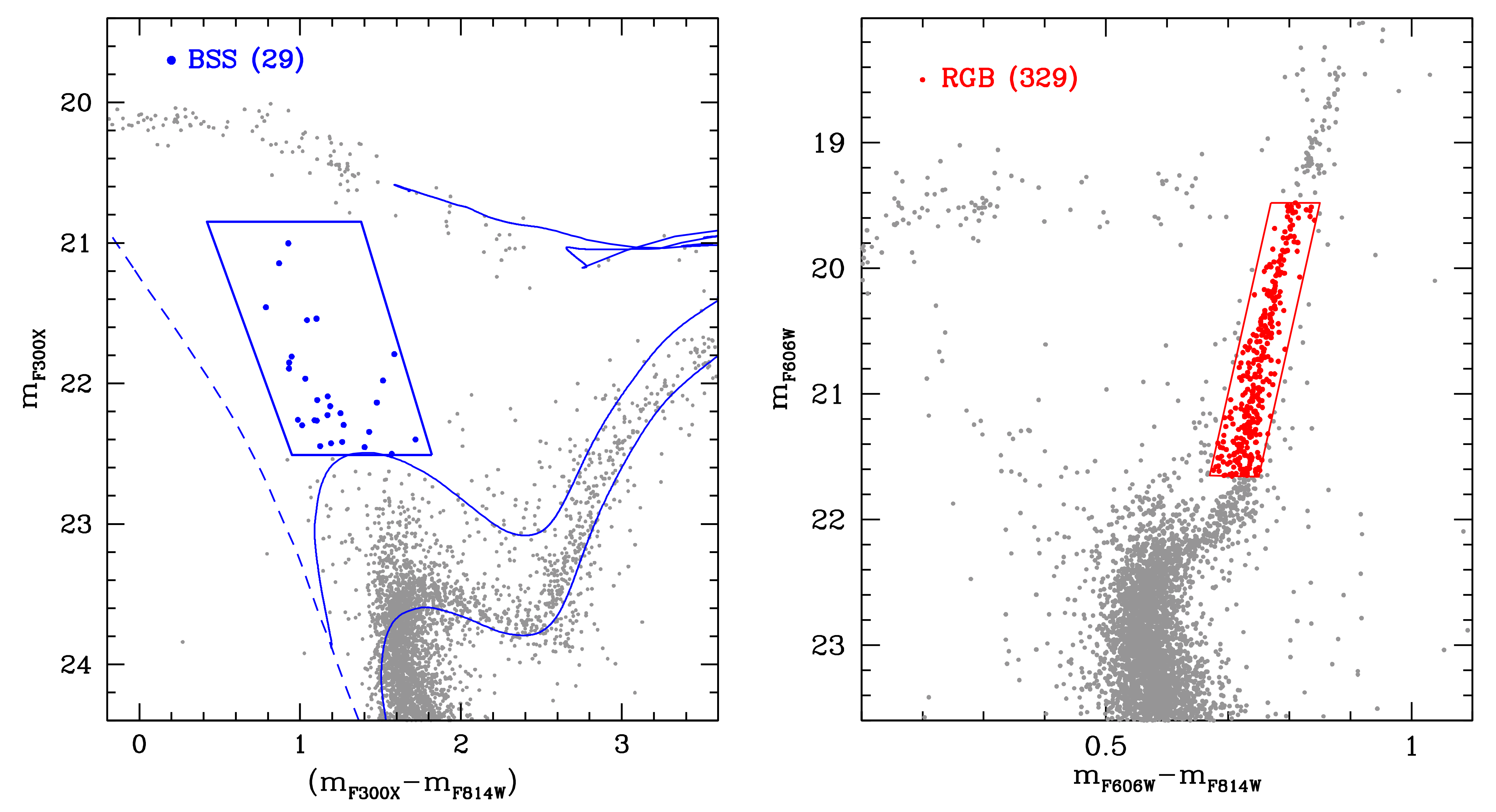}
         \caption{Selections of the BSS and RGB samples. The left panel shows the near-UV CMD  of 
         NGC 1754 within the half-mass radius ($r<r_h=13\arcsec$; grey dots), with the selected BSS sample shown with blue circles and the BSS selection box marked in blue. The rightmost solid line is the evolutionary track for a star of $0.78 M_{\odot}$ (which is the MS turn-off mass for an age $t=12.8$ Gyr). The leftmost solid line is the evolutionary track of a $0.98 M_{\odot}$ star. The dashed line is the BASTI isochrone corresponding to a very young age (30 Myr), representing the cluster ZAMS. The right panel shows the optical CMD of NGC 1754 for $r<r_h=13\arcsec$ (grey dots), with the selected RGB sample shown as red circles and the RGB selection box also marked in red. }
         \label{boxes}
\end{figure*}

\subsection{Star count density profile and structural parameters}
\label{sec:densprof}
Once determined the cluster gravitational center, we constructed the density profile of the system from resolved star counts
\citep[e.g.,][]{miocchi+2013, lanzoni+2019, raso+2020, giusti+2024b}. For this operation, we used only stars with  $m_{\rm F814W} < 22$, hence only those that are brighter than the MS turn-off and therefore have approximately the same mass (the only exceptions are BSSs, which however are so few in number that their contribution to the sample used for the determination of the density profile is negligible). We divided the WFC3 field of view into concentric rings ranging from sub-arcsecond scales up to about 120\arcsec, with each ring further subdivided into three or four sectors.
For every radial bin, we counted the number of stars in each sector and we calculated the star density as the average of the resulting values divided by the sampled area, and the error as their standard deviation. The observed star count density profile obtained by this procedure is marked by the empty circles in Fig.\ref{profili}. This profile features a flat inner portion with a density of approximately $\log\Sigma_{*}\sim1.2$ up to $\sim$ 1\arcsec, a subsequent radial decrease and a final plateau. We averaged the plateau points (the last two points of the profile) to obtain the value of the field star density ($\log\Sigma_{*}\sim-1.05$), and then we subtracted this value from each bin to obtain the field-decontaminated cluster profile (see the solid circles in Fig.\ref{profili}).

Using stars of approximately the same mass allowed us to compare the decontaminated profile with the family of spherical, isotropic, single-mass King models \citep{king+1966}. This comparison was performed using the MCMC method with flat priors for the fitting parameters (central density, concentration
parameter $c$, and core radius $r_c$). The likelihood was based on a  $\chi^2$ statistic. The final best-fit model is represented by the red line in Fig. \ref{profili}.  The residuals in the bottom panel of the figure confirm that a (flat-core) King model accurately reproduces the profile in the central region, effectively ruling out the possibility that the cluster has already undergone core collapse.

The structural parameters obtained by the profile fit procedure are: concentration parameter c =$1.47$, core radius $r_c = 3.5\arcsec$, half-mass radius $r_h = 13\arcsec$, and tidal radius $r_t = 108\arcsec$. Assuming a distance of 49.6 kpc for the LMC \citep{pietrzynski+2019}, we obtain $r_c =$  0.84 pc, $r_h =$ 3.13 pc,  and $r_t = $25.98 pc (see Table \ref{table:1}). This confirms the cluster's compact nature, as anticipated by previous studies. Notably, our estimate of the core radius aligns well, within the uncertainties, with the literature value of $r_c$ = 3.61 $\pm 0.49$\arcsec \citep{mackey+2003}, which was obtained using a different method that involved fitting the brightness profile with an Elson-Fall-Freeman model \citep{elson+1989}.
 
\subsection{Measuring the dynamical age}
\label{BSS} 
Given the cluster's old age and compact structure, a very high dynamical age is anticipated, according to what was mentioned in the Introduction. A parameter that is often used for the characterization of the dynamical status of a cluster is the current central relaxation time, $t_{rc}$.
According to \citet{djorgovski+1993}, this parameter is defined as follows:
\begin{equation}
    t_{rc}=1.491\times10^7 \times \frac{k}{\ln(0.4N_{*}) m_{*}}\rho_{M,0}^{1/2} r_c^3,
\end{equation}
where $k$ is a constant with an approximate value of 0.5592, $N_*=M_{cl}/m_*$ represents the total number of stars in the cluster, where $M_{cl}$ is the total cluster mass and $m_{*}=0.3 M_\odot$ the average stellar mass. The parameter $\rho_{M,0}$ denotes the central mass density in units of $M_\odot$ pc$^{-3}$, $r_c$ is the core radius in parsecs.  We adopted the cluster mass from \citet{mackey+2003}, originally estimated using a mass-to-light ratio of $M/L_V = 3.36$, and rescaled it to $M/L_{V}=2$ to align with the assumptions of \citet{ferraro+2019}. In addition we measured $\rho_{M,0}$ from equations (4), (5), (6) of \citet{djorgovski+1993}, adopting the concentration parameter $c$ and the core radius $r_c$ determined from the fit to the star density profile (see Sect. \ref{sec:densprof}), and the central surface brightness $\mu_{555}(0) = 17.48$ quoted in \citet{mackey+2003}. The result obtained for NGC 1754 is $\log(t_{rc}/{\rm yr}) = 7.969$ (see Table \ref{table:1}). This value is 2 orders of magnitude smaller than the measured chronological age, thus indicating that the cluster core has been significantly affected by two-body relaxation and pointing to an
advanced dynamical age for the cluster. Indeed, \citet{ferraro+2018a} used the parameter $N_{\rm relax}$ (defined as the ratio of the chronological age of a cluster to its central relaxation time) to quantify this information. Of course higher values indicate clusters that are more dynamically evolved. In the case of NGC 1754, given a chronological age of t= 12.8 Gyr,  $N_{relax}$ corresponds to a relatively large value of 137 (see Table \ref{table:1}). 

However, it is important to emphasize that $t_{rc}$ provides just a qualitative indication of the level of dynamical evolution reached by the cluster, since it is evaluated from the present-day structural parameters of the system (which, instead, are known to vary with the dynamical history) and it is estimated through a simplified analytical expression derived under assumptions (such as spherical symmetry, orbital isotropy, and non rotating internal kinematics) that are challenged by recent kinematic investigations (see, e.g.,  \citealt{fabricius+2014, watkins+2015, bellini+2017, ferraro+2018b, kamann+2018, lanzoni+2018a, lanzoni+2018b, leanza+2022}).

Hence, to verify this result and more accurately quantify the dynamical age of NGC 1754, we applied the dynamical clock method \citep{ferraro+2018a, ferraro+2020, ferraro+2023}. As described in Sect. \ref{sec:intro}, this method evaluates the system's dynamical state by analyzing the degree of central segregation of BSSs compared to a lighter-mass reference population.

\begin{table}[h!]
\caption{Main properties of NGC 1754.}
\centering
\begin{tabular}{l l}
\hline
\hline
Parameter & Estimated value \\
\hline
Center of gravity & $\alpha_{J2000}$ = $04^h54^m18^s.66$\\
 & $\delta_{J2000}$ = $-70^{\circ}26' 31.03''$ \\
Age &  $t=12.8^{+0.4}_{-0.4}$ Gyr \\
King concentration & $c=1.47^{+0.12}_{-0.11}$\\
Central dimensionless potential & $W_0=6.8^{+0.4}_{-0.4}$ \\
Core radius & $r_c = 3.5^{+0.5}_{-0.5}$ arcsec (0.84 pc)\\
Half mass radius & $r_h = 13^{+2}_{-1}$ arcsec (3.13 pc)\\
Tidal radius & $r_t = 108^{+22}_{-16}$ arcsec (25.98 pc)\\
Central relaxation time  & $\log(t_{rc}/{\rm yr}) =$ 7.969\\
Age/Central relaxation time & $N_{\rm relax} = 137$\\
\hline
\label{table:1}
\end{tabular}
\vspace{0.15cm} 

\end{table}

\begin{figure*}
     \centering
         \includegraphics[scale = 0.40]{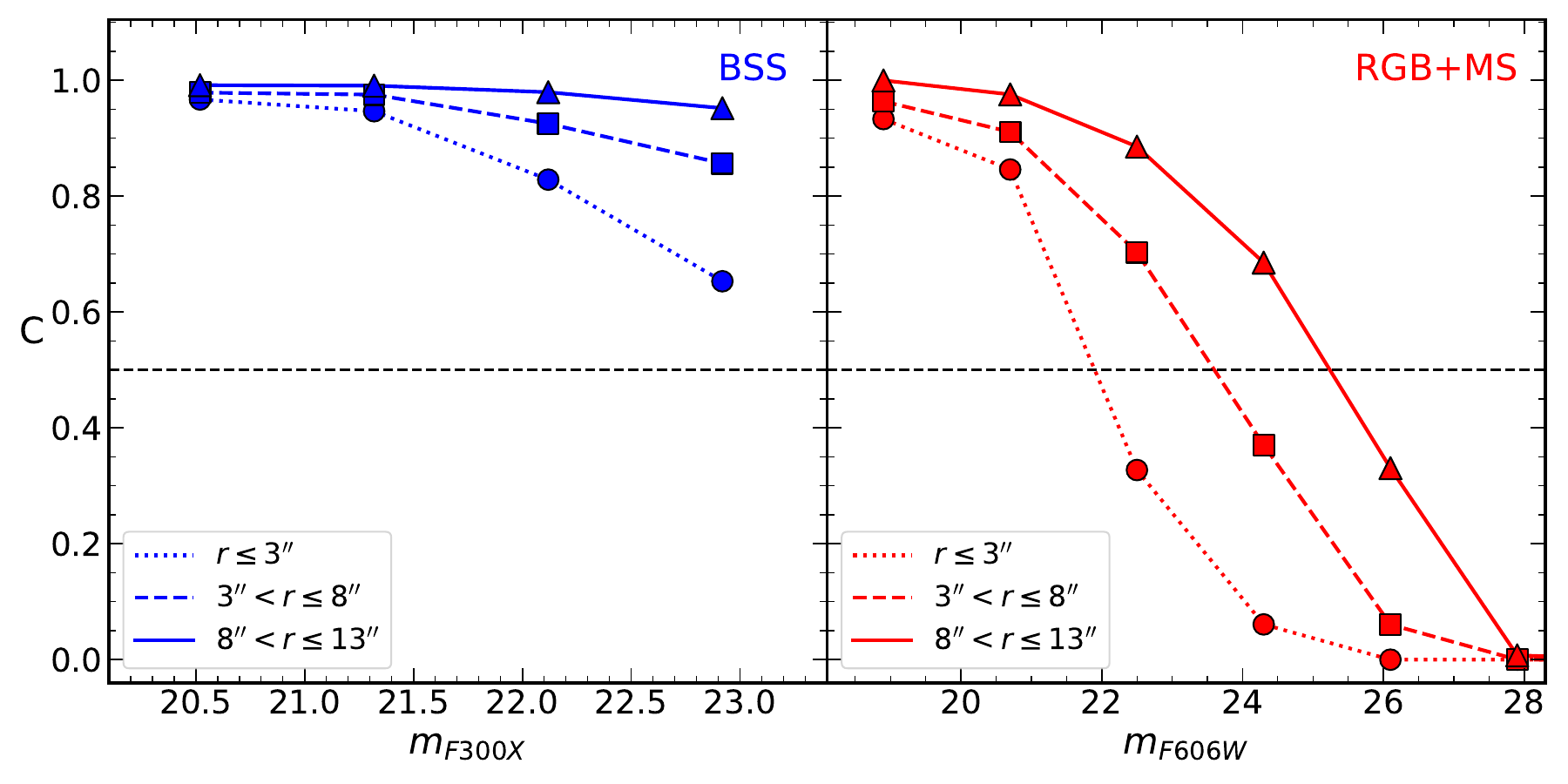}
         \caption{Completeness curves of the BSS and RGB+MS populations as a function of the $m_{\rm F300X}$ and the
           $m_{\rm F606W}$ magnitudes (left and right panels, respectively). The dotted, dashed, and solid lines correspond respectively to the innermost ($r<3\arcsec$), intermediate ($3\arcsec <r< 8\arcsec$), and outermost ($r>8\arcsec$) radial bins considered in the analysis. 
           }
         \label{completezza}
\end{figure*}

\begin{figure*}
     \centering
         \includegraphics[scale = 0.20]{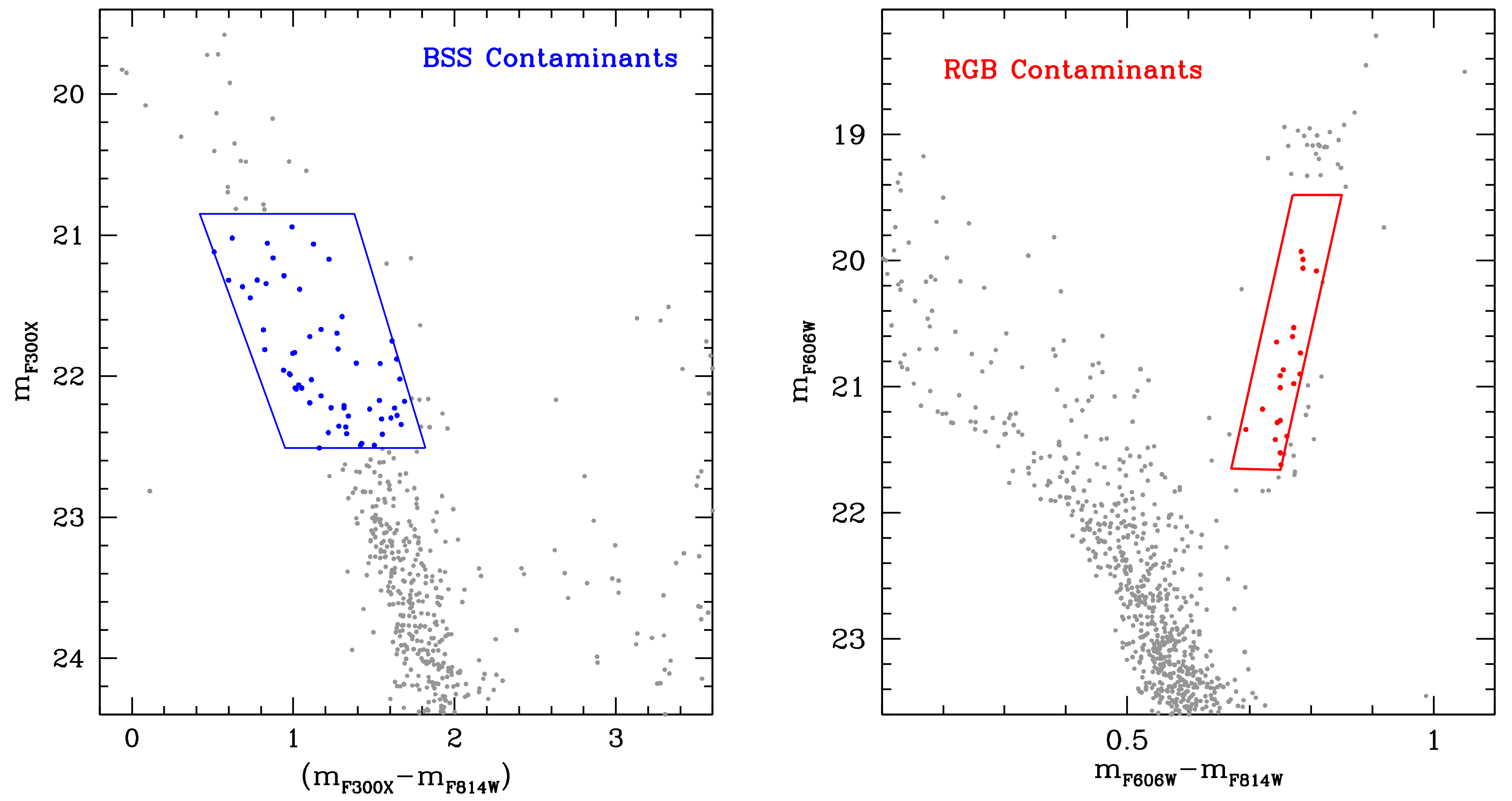}
         \caption{Estimating the LMC field contamination. The BSS and RGB selection boxes shown in Fig.\ref{boxes} are over-plotted, respectively, to the UV (left panel) and optical (right panel) CMDs observed beyond the cluster tidal radius ($r>108\arcsec$).}
         \label{field}
\end{figure*}

The first step is to select a sample of bright BSSs within the cluster’s half-mass radius (according to \citealt{alessandrini+2016,lanzoni+2016} definition). Because BSSs are hot stars, the selection is expected to be most effective in an ultraviolet CMD \citep{ferraro+2003, raso+2017}, here constructed with the F300X filter. Fig.\ref{boxes} shows the ($m_{\rm F300X}, m_{\rm F300X}- m_{\rm F814W}$) CMD within $r_h = $13\arcsec (see Sect.\ref{sec:densprof}) and the box chosen for the BSS selection. The box was defined as follows (see \citealp{ferraro+2018a, ferraro+2019, ferraro+2023, giusti+2024b}), using BASTI isochrones and tracks \citep{pietrinferni+2021} computed for [Fe/H]$= -1.45$, [$\alpha$/Fe]$= +0.4$, $Y = 0.25$, and shifted in the observed CMD by assuming the distance modulus and the color excess determined in Sect.\ref{sec:age}. The left side was constructed following approximately the slope of the zero-age MS (ZAMS, shown as a dashed line in Fig.\ref{boxes}). The upper and right-hand sides were constructed trying to exclude stellar populations belonging to different evolutionary phases, such as HB stars and possible blends with RGB stars. The location of the bottom side is instead set with the aim of selecting the most massive BSSs in the sample, thus maximizing the sensitivity of the $A^+_{rh}$ parameter, since the most massive stars are those experiencing the strongest effect from dynamical friction. In agreement with the selections adopted in previous papers \citep{ferraro+2018a, ferraro+2019, giusti+2024b}, here we included in the sample BSSs with masses exceeding the MS turn-off mass by at least 0.2 $M_{\odot}$. Since the MS turn-off mass of the adopted BASTI isochrone at the cluster age ($t=12.8$ Gyr; see Sect. \ref{sec:age}) is 0.78 $M_{\odot}$, the lower boundary of the BSS selection box is marked by the evolutionary track of a 0.98 $M_{\odot}$ star. The final selection resulted in a sample of 29 BSSs. The right panel of Fig. \ref{boxes} shows the selection of the reference (RGB) sample in the optical ($m_{\rm F606W}, m_{\rm F606W}-m_{\rm F814W}$) CMD, consisting of 329 RGB stars. Small changes in the shape of the boxes lead to negligible changes in the result.

\begin{table*}
\caption{The BSS and RGB samples.}
\renewcommand*{\arraystretch}{2.0}
\centering 
\begin{tabular}{c c c c c}
\hline
& N$_{\rm TOT}$ & N ($r\leq 3\arcsec$) & N ($3\arcsec< r \leq 8\arcsec$) & N ($8\arcsec \leq r \leq 13\arcsec$) \\
\hline
BSS & 33 (8) & 15 (0) & 11 (2) & 7 (6)\\
RGB & 378 (3) & 80 (0) & 199 (1) & 99 (2) \\
\hline
\label{table:2}
\end{tabular}
\tablefoot{Completeness-corrected samples of BSSs (top row) and RGB stars (bottom row). N$_{\rm TOT}$ represents the
  total number of stars, while the last three columns show the estimated 
  numbers of stars in the three radial bins. Numbers in brackets indicated the estimated field
  contaminants. }
\end{table*}

Once the two samples were chosen, we had to take into account their level of photometric completeness. To do this, we performed artificial star experiments using the prescriptions discussed in 
\citet[see also
\citealt{cadelano+2020a, giusti+2024b}]{dalessandro+2015}. As a first step, both for the cluster's principal sequences (MS and RGB) and for the BSS sequence we constructed two mean ridge lines, in the ($m_{\rm F606W}, m_{\rm F606W}-m_{\rm F814W}$) CMD and in the ($m_{\rm F606W}, m_{\rm F300X}-m_{\rm F606W}$) CMD. We generated a set of artificial stars with an input $m_{\rm F606W}$ magnitude derived from a luminosity function designed to mimic the observed one for both the BSS and MS+RGB samples. Each artificial star was assigned a ($m_{\rm F606W}$-$m_{\rm F814W}$) and ($m_{\rm F300X}$-$m_{\rm F606W}$) color value, determined based on the corresponding mean ridge line.  We then added these stars to the WFC3 images using DAOPHOT/ADDSTAR, dividing the field of view into cells with sides equal to ten times the full-width at half-maximum of the PSF and adding one star in each cell, so as to avoid crowding effects. On these modified images we then performed the same data reduction procedure described in Sect. \ref{sec:dataanalysis}, using the PSF models already found. We performed this procedure only in the WFC3 UVIS1 chip, since the analyzed BSS and RGB samples have been selected just within the half-mass radius (13\arcsec). At the end of the procedure, we obtained a catalog of more than one million simulated sources for both the BSS and the MS+RGB samples, containing for each artificial star the information given in input (i.e., the magnitudes in the three filters and the coordinates) and the corresponding output information obtained after the data reduction. Using these catalogs we constructed completeness curves, with the completeness parameter $C$ being defined as the ratio between the number of output and the number of input stars ($C=N_{\rm output} /N_{\rm input}$). Figure \ref{completezza} shows the completeness curves relative to three different radial bins within 13\arcsec from the center, obtained for BSSs (as a function of the $m_{\rm F300X}$ magnitude; left panel) and for the MS+RGB sample (in the $m_{\rm F606W}$ filter; right panel). Using these curves, we assigned a completeness value to each of the 29 BSSs and each of the 329 RGB stars according to their magnitude and radial distance from the cluster center. 

The next key step is the evaluation of the level of  field contamination in the considered BSS and RGB samples. To this end, we took advantage of the fact that the WFC3 field of view is larger than the estimated cluster tidal radius ($r_t=108\arcsec$), which implies that the stellar population in the most external regions (at $r>r_t$) can reasonably be assumed to be representative of the LMC field. This region covers an area of approximately 4000 arcsec$^2$, thus allowing a reasonable evaluation of the density of the LMC field population in the proximity of NGC 1754. This approach, compared to the use of the ACS data set, allowed us to work directly in the $m_{\rm F300X},m_{\rm F300X}- m_{\rm F814W}$ plane where the BSS have been selected, whereas in the ACS data set the F300X filter is not present. Thus, the selection boxes have been drawn in the field CMD (see Fig. \ref{field}) and the number of stars falling in each of them (corresponding to potential contaminants to the BSS and RGB populations) has been counted. The ratio between these number counts and the sampled area gives the value of the expected field contamination density in each population, which turns to be $0.015$ and $0.005$ contaminant star per square arcsecond for the BSS and the RGB population, respectively. Considering that the portion of cluster within the half-mass radius ($r_h=13\arcsec$) subtends approximately 531 sq.arcsec on the sky, we estimate a total of 8 and 2.7 (in the following we assumed 3) contaminants in the BSS and RGB samples, respectively. However, for a more precise field decontamination we evaluated the number of field interlopers in three distinct radial bins, the same used to estimate the completeness curves (Fig. \ref{completezza}). To this end, we multiplied the field densities in the BSS and RGB boxes by the area of each annulus, finding the values listed in brackets in Table \ref{table:2}.

\begin{figure}
         \includegraphics[scale = 0.68]{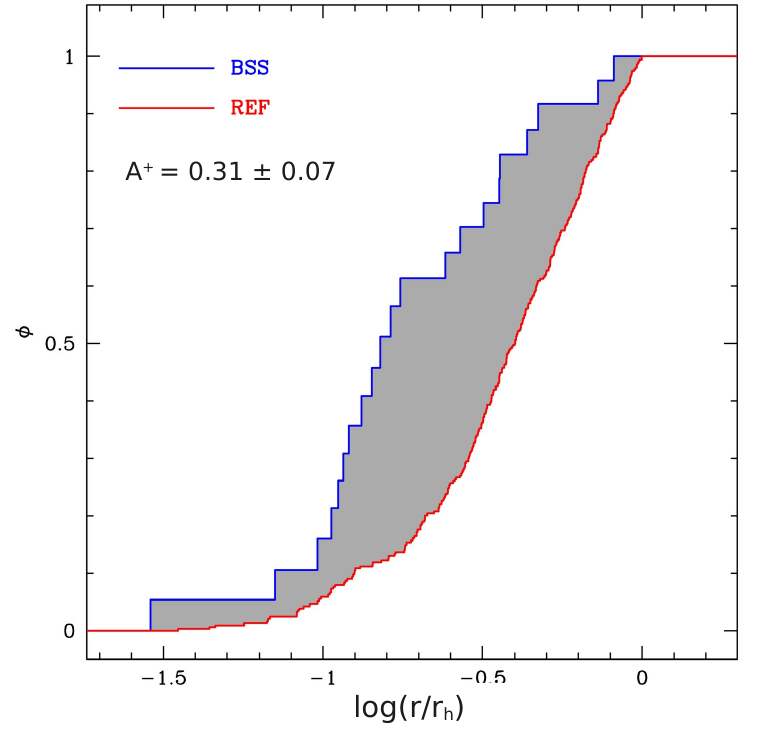}
         \caption{The normalized cumulative radial distributions of BSSs (solid blue line) and RGB stars (solid red line) from one of the random realizations used to estimate the $A^+_{rh}$ parameter. The $A^+_{rh}$ value represents the area enclosed between the two curves.}
         \label{apiu}
\end{figure}

To measure the $A^+_{rh}$ parameter, we constructed the cumulative radial distributions of both the BSS and RGB samples after proper correction for incompleteness effects by weighting each star by the inverse of its assigned completeness parameter $C$. 
Additionally, to account for field contamination we repeated the calculation of $A^+_{rh}$ 1000 times, each time randomly removing the estimated number of contaminating field stars. In each individual iteration, the value of $A^+_{rh}$ was measured as the area enclosed between the cumulative radial distribution of the BSS sample and that of the RGB sample. Figure \ref{apiu} presents a random realization of the
cumulative radial distributions of BSSs (blue curve) and RGB stars (in
red), with the grey-shaded area indicating the $A^+_{rh}$ parameter. The final $A^+_{rh}$ value was obtained as the mean of the 1000 individual measurements, resulting in $<A^+_{rh}>=0.31$, which indicates an old dynamical age for NGC 1754. While the formal 1-$\sigma$ uncertainty on this value is only 0.01, to obtain a more realistic estimate also accounting for the relatively small number of stars, the error associated to this final value was measured using a jackknife bootstrapping technique. Specifically, for a sample of $N$ BSSs, the measurement of the area enclosed within the two cumulative distributions was repeated $N$ times, each time randomly removing one star from the sample. The error is then defined as $\sigma_{A^+_{rh}} = \sigma_{\rm distr} \times \sqrt{(N-1)}$, where $\sigma_{\rm distr}$ is
the standard deviation of the $N$ estimates of $A^+_{rh}$. In our case $\sigma_{A^+_{rh}} = 0.07$.

\begin{figure*}
     \centering
         \includegraphics[scale = 0.22]{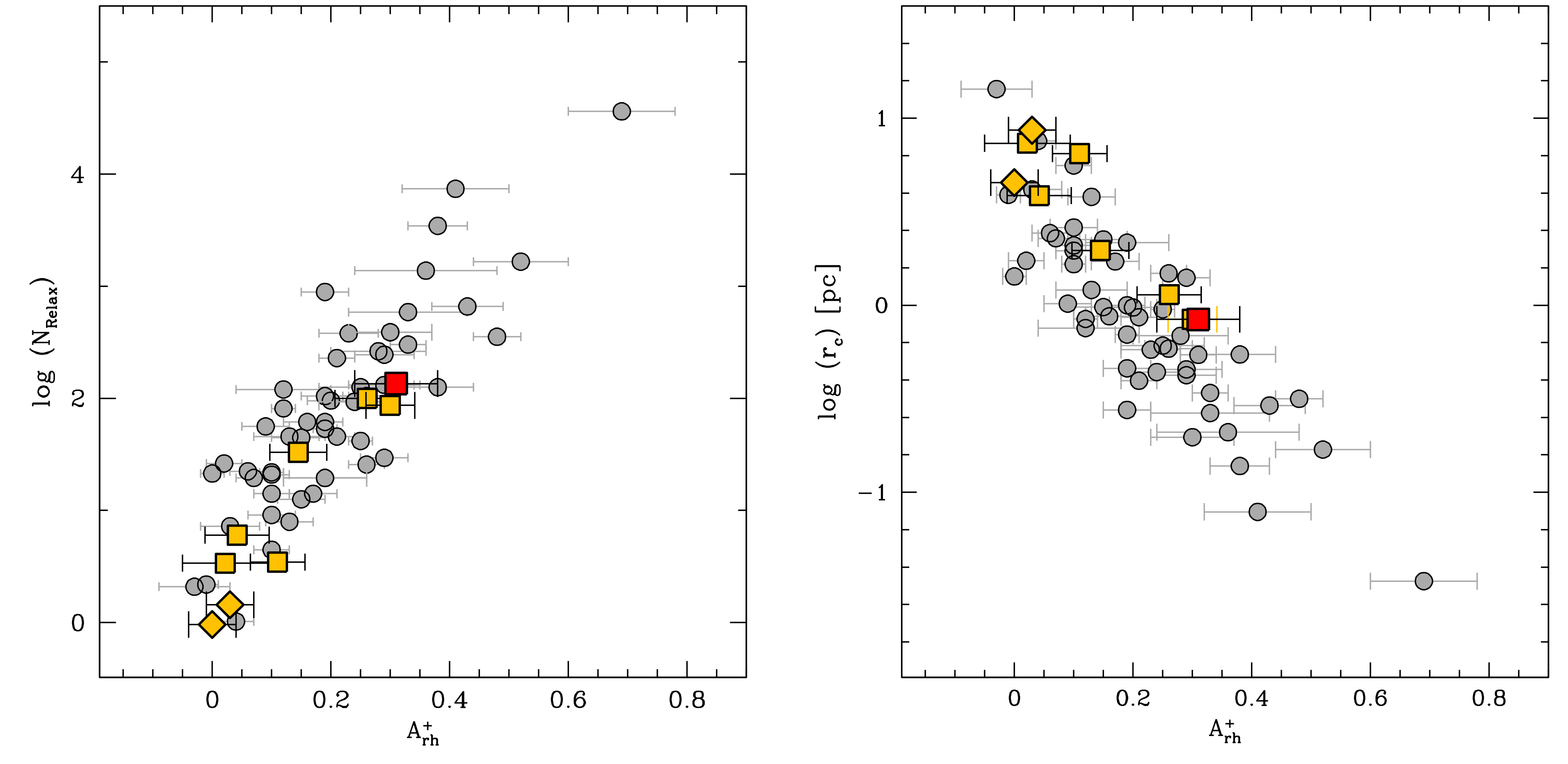}
         \caption{The left panel illustrates the relation between $N_{\rm relax}$ and $A^+_{rh}$, while the right panel the one between $r_c$ and $A^+_{rh}$ for star clusters  analyzed using the dynamical clock method. NGC 1754 is represented by a large red square, while the 52 Galactic GCs discussed by \citet{ferraro+2023} are shown as gray circles. The six old LMC clusters from \citet{ferraro+2019} and \citet{giusti+2024b} are plotted as yellow squares, and the two young SMC clusters from \citet{dresbach+2022} are marked with yellow diamonds.}
         \label{conundrum}
\end{figure*}

\section{Discussion and conclusions}
\label{disc}
The analysis presented in this paper confirms a very old chronological age for NGC 1754 ($t=12.8\pm0.4$ Gyr), fully comparable with that measured both in the other old LMC clusters (see \citealp{wagner+2017, giusti+2024a}) and in the oldest MW GCs (see, e.g., Fig. 2 in \citealt{ferraro+2019}). This result further consolidates the indication that the beginning of the GC formation process was contemporaneous in the MW and the LMC. Indeed, this is a relevant point since the relative distance between the two galaxies was much larger at the epoch of GC formation than today (as suggested by their orbit reconstruction; see \citealt{kalli+2013}) and they offered independent and very different cluster formation environments in terms of total mass, metallicity, etc. Hence, this evidence points toward a GC formation process starting at the same cosmic time independently of the characteristics of the hosting galaxy.

Concerning the dynamical characterization of the cluster, we estimate that it is already in an advanced dynamical stage, and we can now compare the results obtained for NGC 1754 with those published in the literature for other clusters. A first conclusion can be drawn by examining the plane that relates the measured values of $A^+_{rh}$ to those obtained for the parameter $N_{\rm relax}$, which is defined as the ratio between the chronological age of a cluster and its central relaxation time $t_{rc}$ \citep{ferraro+2018a}. This is shown in the left panel of Figure \ref{conundrum}, for NGC 1754 (red square), the 52 Galactic GCs discussed in \citet{ferraro+2023} (gray circles),  
along with eight Magellanic Cloud systems: six old clusters in the LMC (yellow squares; \citealt{ferraro+2019, giusti+2024b}) and two young ones in the SMC (yellow diamonds; \citealt{dresbach+2022}).   
In the figure, the distribution of the MW and LMC clusters defines two perfectly superposed, almost diagonal sequences.
The observed trend clearly indicates that both parameters are measuring the same physical phenomenon, the dynamical aging of star clusters. However, as discussed in \citet{ferraro+2023}, the $A^+_{rh}$ parameter offers a series of advantages with respect to the current central relaxation time. In fact, as quoted in the previous section, the present-day central relaxation time is estimated under quite strong simplifying assumptions and it is based on the current structure of the cluster, thus only providing an ``instantaneous shot'' of the system dynamical status. 
Hence, as a general rule, the value of $N_{\rm relax}$ can be considered a first-guess proxy of the dynamical age of the cluster but it surely cannot be used as tracer of its complete past evolutionary history. In contrast, by measuring the observed concentration of the BSS population, $A^+_{rh}$ provides a direct empirical measurement of the level of mass segregation developed during the entire cluster’s evolution and allows a more detailed analysis and comparison between different clusters.

As can be appreciated, NGC 1754 (red square) nicely nests into the correlation between $A^+_{rh}$ and $N_{\rm relax}$, further extending the distribution of the LMC clusters, since it turns out to be the most dynamically evolved system studied so far in the Magellanic Clouds. Intriguingly, the value of $A^+_{rh}$ measured for NGC 1754 ($A^+_{rh}=0.31$), as well as that of NGC 1835 ($A^+_{rh}=0.30$; see \citealt{giusti+2024a}) are very close to the value that seems to flag the core collapse (CC) event. In fact, as discussed in \citet{ferraro+2023}, $A^+_{rh}=0.30$  can be considered a sort of reference value for CC, meaning that the proximity of $A^+_{rh}$ to this value is a likely indication of the imminence or the occurrence of the CC event. In fact, all the 7 MW GCs classified as CC (together with NGC 362, which is suspected to be a post-CC system; see the Harris catalog and the discussion in \citealt{dalessandro+2013b}) have $A^+_{rh}\ge 0.29$. 

In the right panel of Fig. \ref{conundrum}, we show the comparison between the values of $A^+_{rh}$ and the core radius for the aforementioned clusters. In this case, an anti-correlation is observed, with NGC 1754 (large red square) well aligning within this trend. The result is fully compatible with what expected from the natural dynamical evolution of star clusters, namely that the cluster structure becomes more compact (i.e., $r_c$ decreases) as the dynamical age increases. As mentioned in the Introduction, 
although this trend is qualitatively expected for all dynamically-active stellar systems, the close match between the results observed in the MW and in the Magellanic Clouds is important and not entirely expected.
In fact, the dynamical evolution is driven by a complex combination of effects associated with a variety of processes (e.g., two-body relaxation, interactions with the Galactic tidal field along the cluster orbit, initial structural and kinematic properties of the cluster and its stellar content; see, e.g.,  \citealt{heggie+2003}). Hence, the striking similarity between the dynamical aging of MW and Magellanic Cloud GCs, which were born and evolved in very different environments, is not obvious and suggests that the properties of the local environment, as the parent galaxy tidal field, play, at most, a secondary role with respect to the internal ones. 
 
For what concerns the size-age conundrum, the results obtained here for NGC 1754 confirm the validity of the scenario proposed in \citet{ferraro+2019}, where the difference in the values of the core radius observed for the old LMC clusters is due to their different dynamical ages, with less compact clusters being less dynamically evolved than the more compact ones. The behavior observed for the youngest clusters (i.e., the lack of young clusters with large core radii) might instead be due to the star formation history of the LMC: an initial star formation burst around 12-13 Gyr ago produced the most massive clusters ($M>10^5 M_\odot$) and, after a 10 Gyr quiescent period, it was followed by a second burst that generated a population of significantly less massive systems ($M<10^5 M_\odot$) close to the galaxy's center \citep{rich+2001, bekki+2004, mazzi+2021}; since the youngest clusters were formed in a region where the galaxy's gravitational forces are very strong, only the most compact ones managed to survive.

\begin{acknowledgements}
This work is part of the project Cosmic-Lab at the Physics and
Astronomy Department ``A. Righi'' of the Bologna University (http://
www.cosmic-lab.eu/ Cosmic-Lab/Home.html). 
\end{acknowledgements}

\end{document}